\documentclass{eas}
\usepackage{graphicx}
\newcommand{\ms}{M$_{\odot}$}

\begin{document}

\title{The chemical evolution of the Milky Way in a cosmological context} 
\author{Nikos Prantzos}\address{Institut d'Astrophysique de Paris, UMR7095 CNRS, Universit\'e Pierre \& Marie Curie}

\begin{abstract}
A short overview is presented of several topics concerning the evolution of the Milky Way (MW) in a cosmological context. In particular,  the metallicity distribution of the MW halo is derived analytically and the halo metallicity and abundance patterns are compared to those of Local Group galaxies. The inside-out formation of the MW disk is supported by the observed evolution of the abundance gradients, while their magnitude suggests that the role of the Galactic bar has been negligible. Finally, the empirical foundations (age-metallicity relation and metallicity distribution) of the evolution of the solar neighborhood, which is the best studied galactic sub-system, have been seriously questioned recently.
\end{abstract}
\maketitle

\section{Introduction}

Studies of the MW in a cosmological framework literally exploded in the past few years. Our Galaxy's properties are far better known than those of any other galaxy (and the situation will tremendously improve with surveys as RAVE, SIM and GAIA), making it an ideal benchmark for tests of galaxy formation theories. I will review here some progress in studies of the MW halo and disk, made mostly after the seminal review of Freeman and Bland-Hawthorn (2002).

\section{The early Milky Way and  hierarchical galaxy formation}

According to the paradigm of hierarchical structure formation, the early phases of a galaxy's evolution are the most complex ones, as they involve multiple mergers of smaller sub-units. In the case of the Milky Way, "chemical signatures" of that period are still around us today, in the form of abundance patterns and metallicity distributions (MD) of long-lived stars.

\subsection{The halo metalliity distribution: outflow vs. subhalo merging}

The MD of Galactic halo field stars (HMD) is rather well known in the metallicity range -2.2$<$[Fe/H]$<$--0.8 (Ryan and Norris 1991), while at lower metallicities its precise form has still to be established by ongoing surveys (Fig. 1 a). Its overall shape is well fitted  by a simple model of GCE as ${{dn}\over{dlogZ}}\propto {{Z}\over{y}} \ e^{-Z/y}$, where $y$ is the $yield$; this function has a maximum for $Z=y$. The HMD peaks at a metallicity [Fe/H]=--1.6 (or [O/H]=--1.1, assuming [O/Fe]$\sim$0.5 for halo stars), pointing to a low yield $y$=1/13 of the corresponding value for the solar neighborhood. Such a low halo yield is "classicaly" (i.e. in the monolithic collapse scenario) interpreted as  due to $outflow$ during halo formation (Hartwick 1976), at the large rate of 8 times the SFR (Prantzos 2003). How can it be understood in the modern framework of hierarchical merging ?

\begin{figure}
\includegraphics[height=5.2cm]{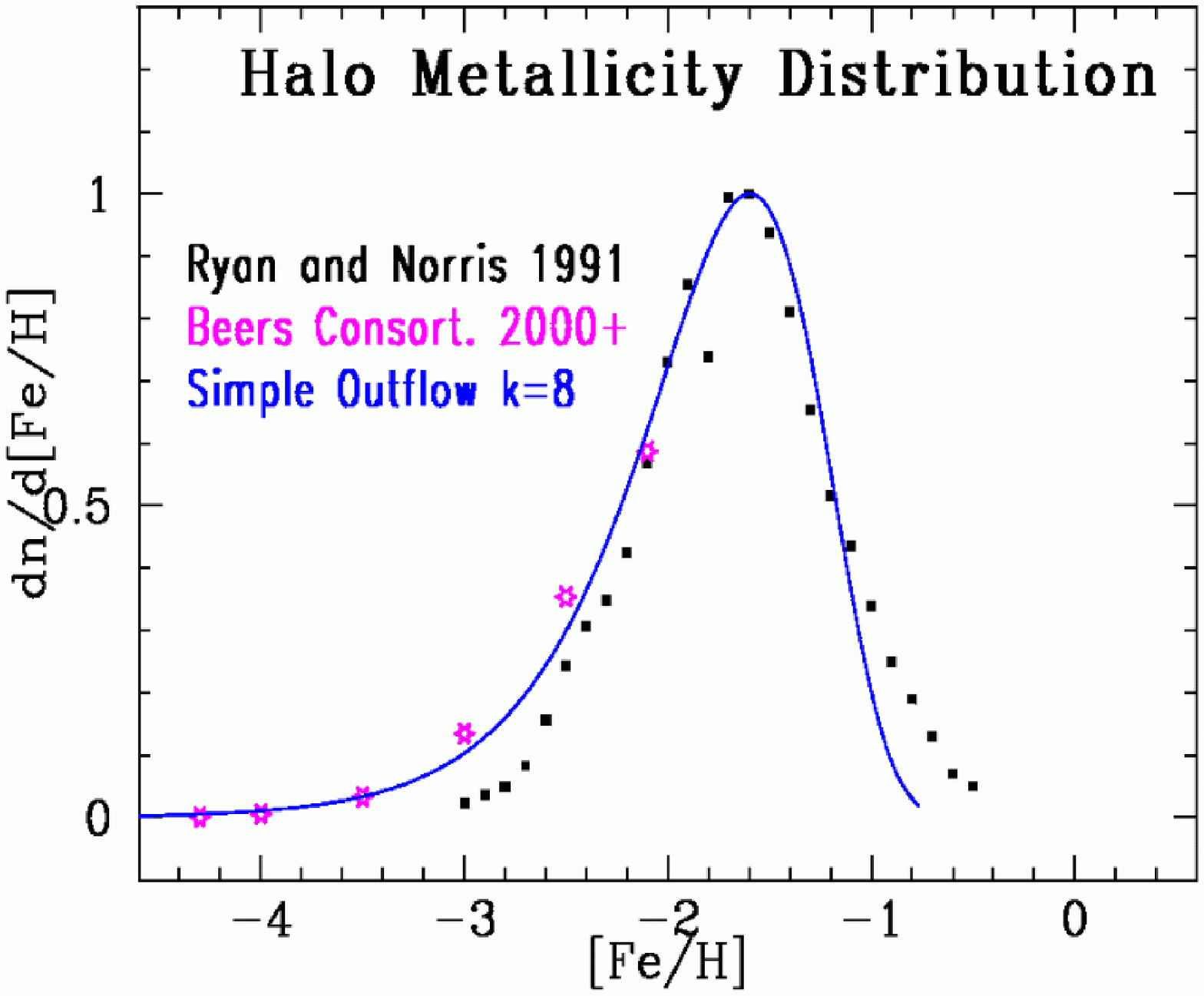}
\qquad
\includegraphics[height=5.2cm]{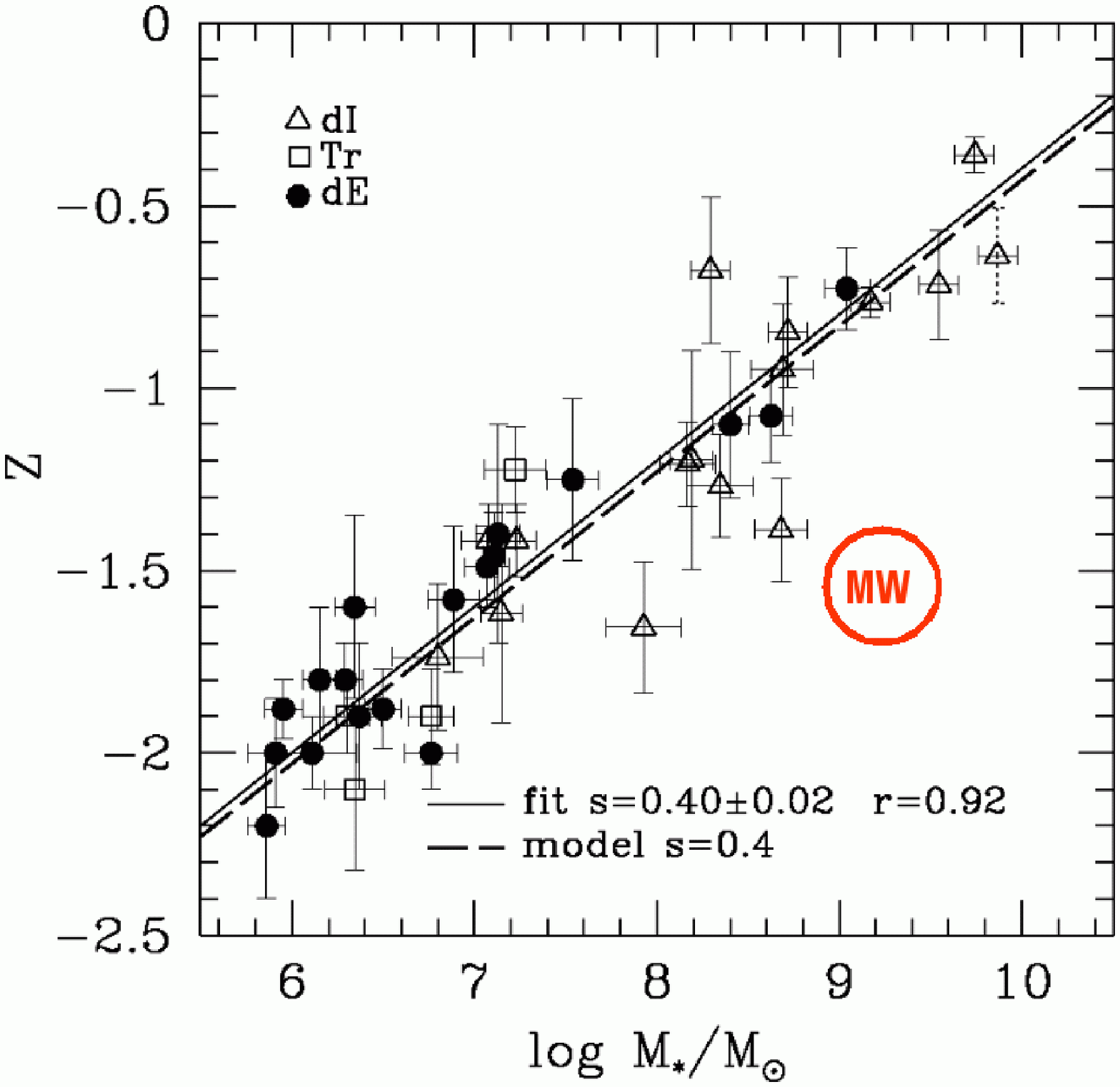}
\caption{{\it Left:} Metallicity distribution of field halo stars from Ryan and Norris (1991, dots), and the ongoing research of Beers and collaborators (T. Beers, private communication, asterisks). The curve is a simple model with outflow rate equal to 8 times the star formation rate. {\it Right:} Stellar metallicity vs stellar mass for nearby galaxies; data and model are from Dekel and Woo (2003). The MW halo, with average metallicity[Fe/H]=--1.6 (see a panel) or [O/H]=--1.1 and estimated mass 2 10$^9$ \ms \ falls below that relationship. }
\label{fig:1}       
\end{figure}

It should be noted that the typical halo metallicity ([Fe/H]=-1.6) is substantially lower, by more than 0.5 dex,  than the corresponding metallicities of nearby  galaxies of similar mass ($M_{Halo}\sim$2 10$^9$ \ms), as can be seen in Fig. 1b. That figure also displays the well known galaxian relationship between stellar mass and stellar  metallicity; most probably, it results from mass loss, which is more important in lower mass galaxies, since the hot supernova ejecta escape more easily their swallow potential well (e.g. Dekel and Silk 1986, Dekel and Woo 2003). 

\begin{figure}
\includegraphics[width=5.9cm]{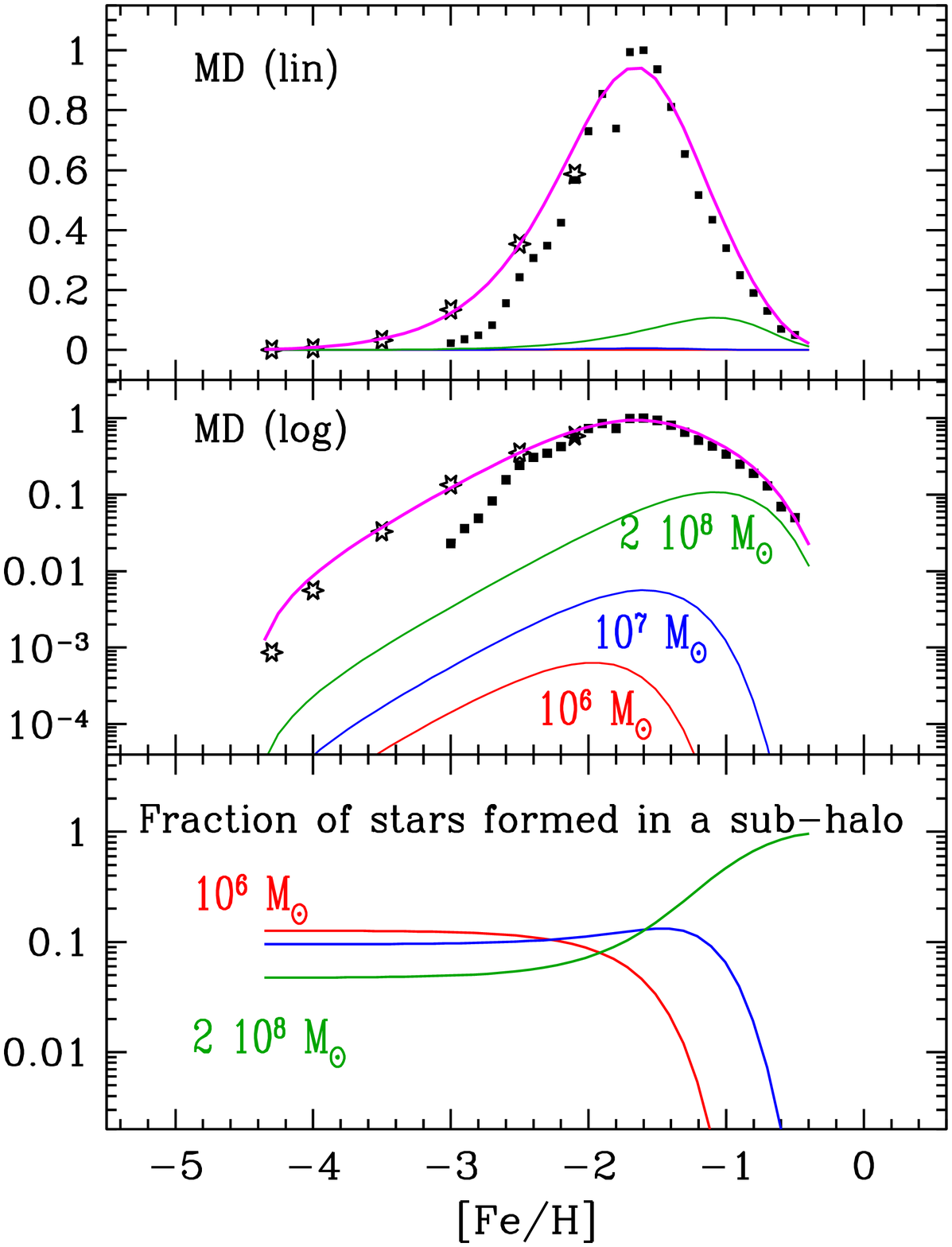}
\qquad
\includegraphics[width=5.9cm]{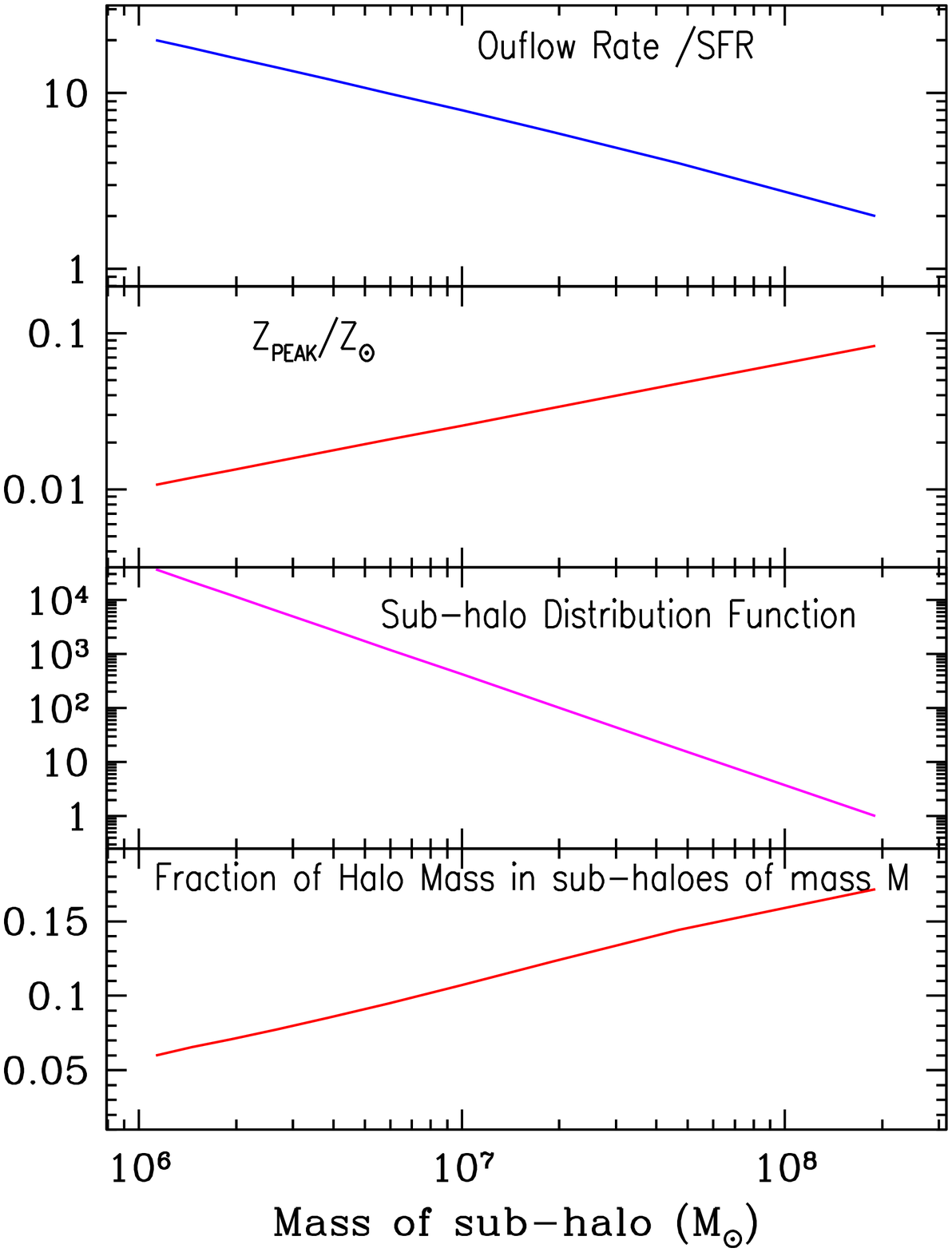}
\caption{{\it Left, top} and {\it middle} panels: Metallicity distribution (in lin and log scales, respectively) of the MW halo, assumed to be composed of a population of smaller units (sub-haloes). The individual MDs of a few sub-haloes, from 10$^6$ \ms \ to 2 10$^8$ \ms, are indicated in both panels (but clearly seen only in the middle), as well as the sum over all haloes (upper curves in both panels, compared to observations). Small sub-haloes contribute the largest fraction of the lowest metalicity stars ({\it bottom left}).  {\it Right panels:} Properties of the sub-haloes as a function of their mass.}
\label{fig:3}       
\end{figure}

Assuming that the MW halo has been assembled from sub-units similar to the low mass galaxies of Fig. 1b, one may interpret the HMD as the sum of  the MD of such low mass galaxies; it is assumed that each one of those galaxies evolved with an appropriate outflow rate and corresponding effective yield $y(M)=(1-R)/(1+k-R)$, where $k(M)$ is the outflow rate in units of the SFR and $R$ is the return mass fraction,  depending on the adopted stellar IMF (see Prantzos 2003). In that case, one has: HMD($Z$) = 1/$M_{Halo}$ $\int {{Z}\over{y(M)}} \ e^{-Z/y(M)} \Phi(M) M dM$, where $\Phi(M)$ the mass function of the sub-units and $y(M)$  the effective yield of each sub-unit (obtained directly from Fig. 1b as $y(M)=Z(M)$, i.e. smaller galaxies suffered heavier mass loss).

The results of such a simple toy-model for the HMD appear in Fig. 2 (left panels). The HMD is extremely well reproduced, down to the lowest metallicities, {\it assuming} $\Phi(M) \propto M^{-2}$; this corresponds to a cumulative mass function $N(>M)\propto M^{-1}$, resulting from high resolution numerical simulations for Milky Way sized dark haloes (Diemand et al. 2006). Low mass satellites (down to 10$^6$ \ms)  contribute most of the low metallicity stars of the MW halo, whereas the high metallicity stars originate in a couple of  massive satellites with M$>$ 10$^8$ \ms \ (Fig. 2, left bottom).

\begin{figure}
\includegraphics[angle=-90,width=0.46\textwidth]{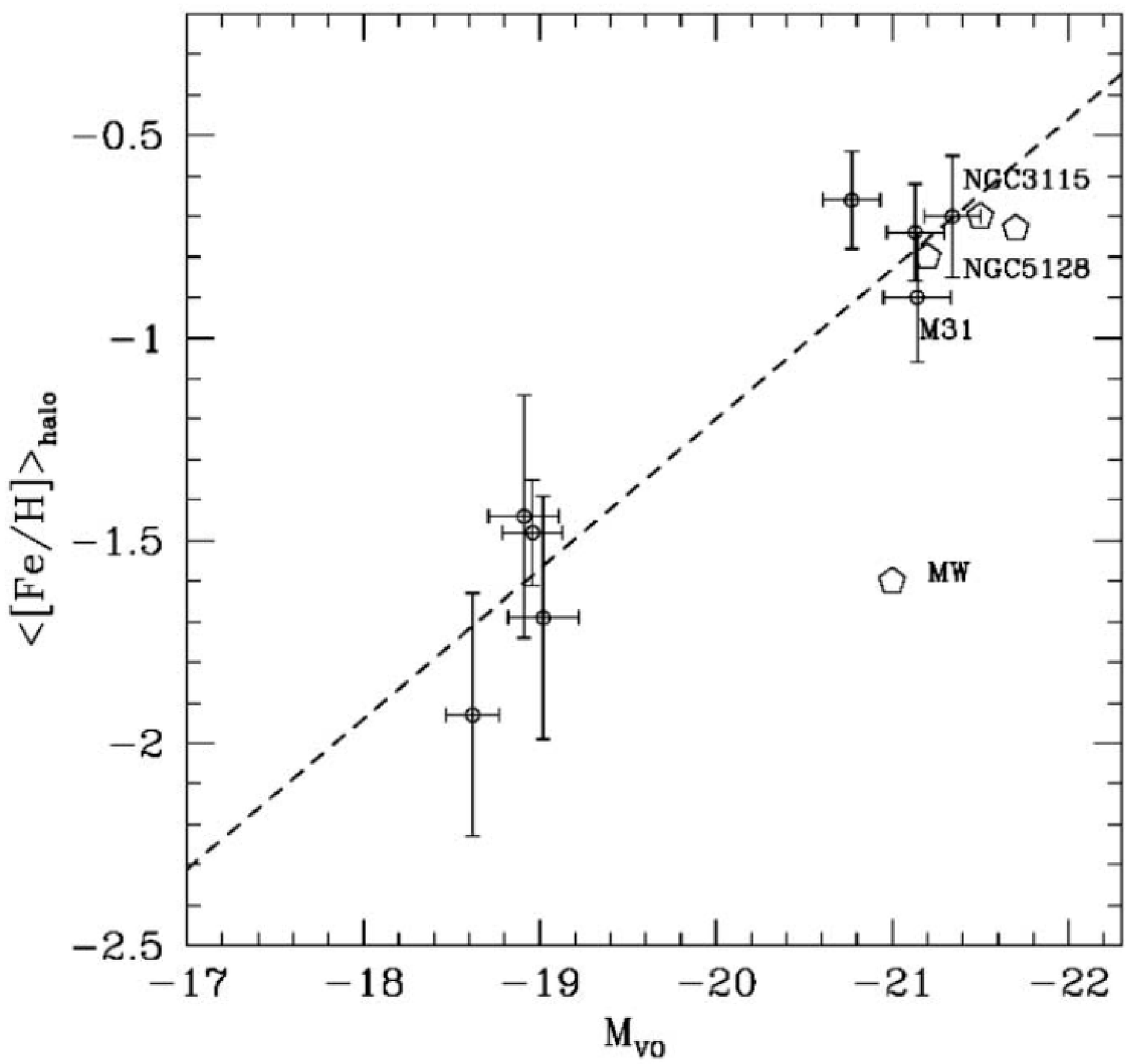}
\qquad
\includegraphics[angle=-90,width=0.46\textwidth]{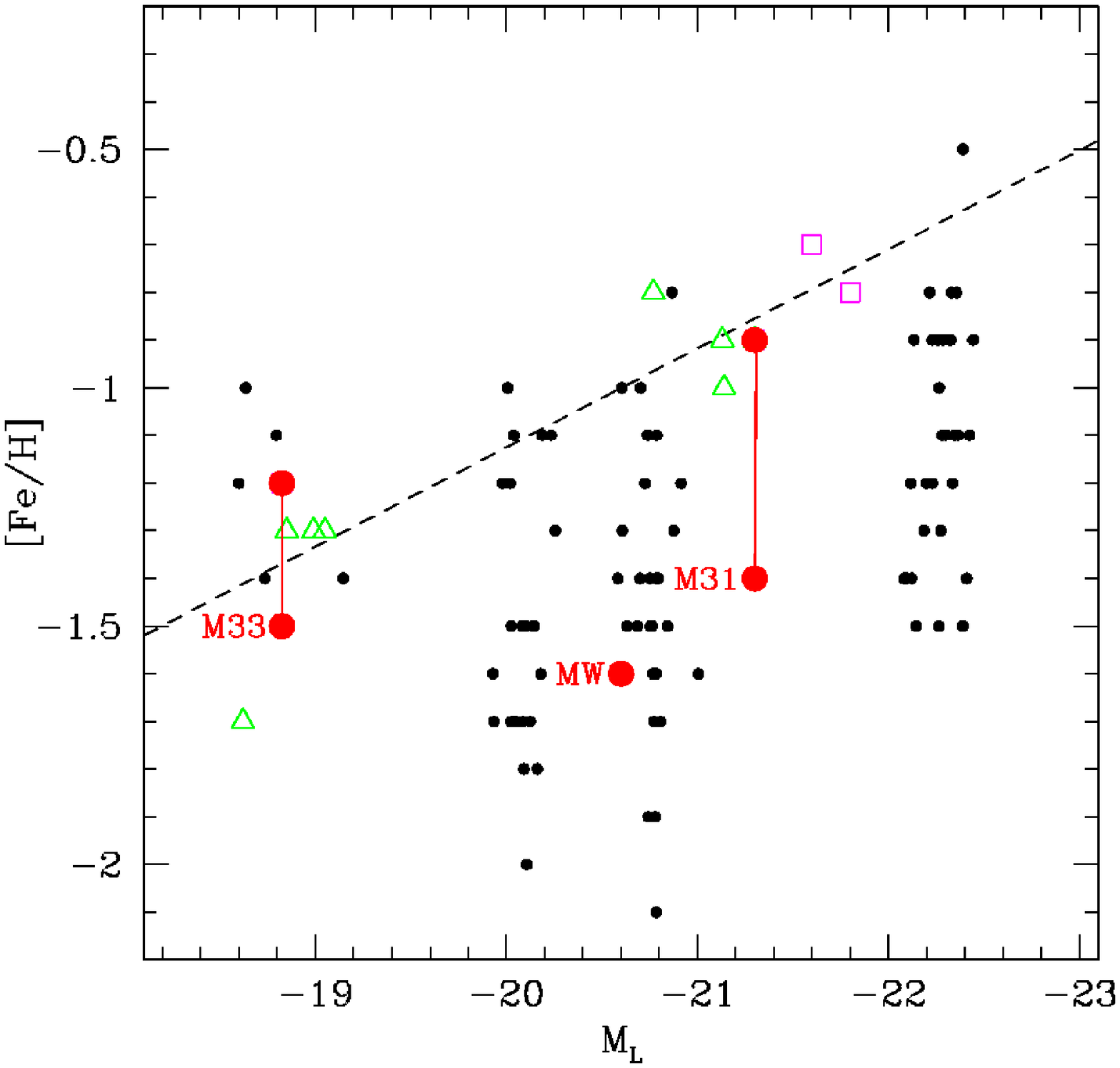}
\caption{{\it Left:} Observed mean metallicity of the halo stellar population vs. luminosity of the parent galaxy; the dashed line is the Dekel-Silk relation for galaxies that suffered mass loss (from Mouhcine et al. 2005). The Milky Way clearly lies below the observed relation. {\it Right:} Open symbols represent same data as in left diagram. Large filled symbols represent spectroscopic values for M33 and M31 (lower points) connected by lines to other literature values (upper points). Now, the MW is quite typical. Small filled symbols represent results of halo models from simulations of Renda et al. (2005). Figure from Chapman et al. (2006).
}
\label{fig:3}       
\end{figure}

Some properties of the sub-haloes  as a function of their mass appear also in Fig. 2 (right panels).  The outflow rate, in units of the corresponding SFR, is $k(M) = (1-R) (y_{True}/y(M) \ - \ 1)$, where $y_{True}$ is the yield in the solar neighborhood ($y_{true}=Z_{peak}$ in Fig. 6b).   If the MW halo were formed in a a potential well as deep as those of comparable mass galaxies, then  the  large outflow rate required to justify the HMD ($k$=8) is puzzling; on the contrary, the HMD is readily understood if the MW halo is formed from a large number of smaller satellites, each one of them having suffered heavy mass loss according to the simple outflow model.

A more physical, but much less ``transparent'', approach consists in deriving the full merger tree of the MW halo and (by using appropriate receipes for SFR and feedback for the sub-haloes) following the chemical evolution through merging/accretion with Monte-Carlo simulations. Recent studies  (Tumlinson 2006, Salvadori et al. 2006) find good agreement with the observed HMD, but it is hard to find (in vue of the many model parameters) what is (are) the key factor(s) determining the final outcome. In any case, {\it ouflow} is crucial in shaping the HMD (albeit in a way different from that envisioned by Hartwick 1976).

\subsection{Metallicities of other nearby haloes}

The stellar population of the MW halo is defined by using both metallicity and kinematics criteria. In the case of nearby galaxies, metallicities are usually defined photometrically, while kinematics is difficult, in general, to determine and spatial criteria are used instead. Mouchine et al. (2005a,b) used observations of the HST to derive photometric metallicities for spatialy selected  field halo red giants in $\sim$80 nearby spirals. They find that (a)  halo metallicity increases with the luminosity of the parent galaxy and (b) the MW halo is undermetallic w.r.t. the haloes of spirals with luminosities comparable to the MW (Fig. 3 a); the latter conclusion corroborates earlier findings of Durell et al. (2001), suggesting that  the field halo population of M31
is substantially more metallic than the MW halo. In view of those results, the MW halo appears rather atypical. 

This picture is, however, at odds with the results of recent studies of M33 (Mc Connahue et al. 2006) and M31 (Chapman et al. 2006), using kinematically selected samples of field halo red giants and spectroscopicaly determined metallicities: the non-rotating halo components of  M33 and M31 have similar metallicities to the MW halo. These findings do not exclude, of course, the existence of more metal rich halo components with different kinematic properties, like those found in the earlier studies. It is possible, however, that such components belong to more massive satellite galaxies accreted later than the most metal poor component. 

From the theoretical point of view, the situation is not clear yet. Early dynamical models of halo formation in CDM framework (Bekki and Chiba 2001) reproduced the peak of the HMD (for reasons related not to outflow but to tidal disruption of mergers), but failed to repoduce the overall shape and smoothness of the HMD (perhaps because of insufficient numerical resolution). In more recent calculations, Font et al. (2006, N-body) find too metal rich haloes w.r.t. the MW, while Read et al. (2006, hydrodynamical) find too metal poor satellite galaxies w.r.t. those in the Local Group. The semi-cosmological chemo-dynamical simulations of Renda et al. (2005) produce a large range of peak metallicities (factor 5-7 variation) for a given halo mass, depending on the number and masses of accreted fragments; the MW halo has a low metallicity for its mass (see Fig. 1 in Renda et al. 2005) but a typical metallicity for a galaxy with the MW luminosity (Fig. 3b).
It should be noted that the problem of {\it identifying} halo particles in the simulations (kinematics, spatial position, or both ?) is not satisfactorily solved yet.

\subsection{Halo abundance patterns and relation to present-day dwarf galaxies}

It has been suggested (e.g. Unavane et al. 1996) that the current paradigm of galaxy formation (through accretion/merging of smaller units)  could be tested by comparing abundance patterns in present day dwarf galaxies and (low metallicity) components of large galaxies, like the MW halo. This concerns, in particular, the $\alpha$/Fe ratio, the only one which has an unambiguous interpretation in the solar neighborhood: both $\alpha$ elements and Fe result solely from SNII  during the MW halo evolution (timescale $<$1 Gyr), while a complementary Fe source (SNIa) operates on longer timescales, reducing the $\alpha$/Fe ratio from $\sim$3 times solar in halo stars to its solar value after $\sim$6 Gyr. Other abundance patterns (involving e.g. r- or s- elements, have much more ambiguous interpretations.

Direct measurements of stellar abundances in dwarf satellite galaxies became feasible only recently, with the advent of 8-10 m. telescopes and efficient spectrographs. An impressive body of data is presented and discussed in Venn et al. (2004). Comparison to the MW halo abundance pattern clearly suggests that the halo was not built from such galaxies, since those satellites have generically low $\alpha$/Fe (implying evolution on timescales $>$1 Gyr, Fig. 4). However, it cannot be excluded that the MW halo accreted such dwarf galaxies {\it early} in their evolution, disrupting them and stopping further evolution involving SNIa.

\begin{figure}
\includegraphics[width=0.47\textwidth]{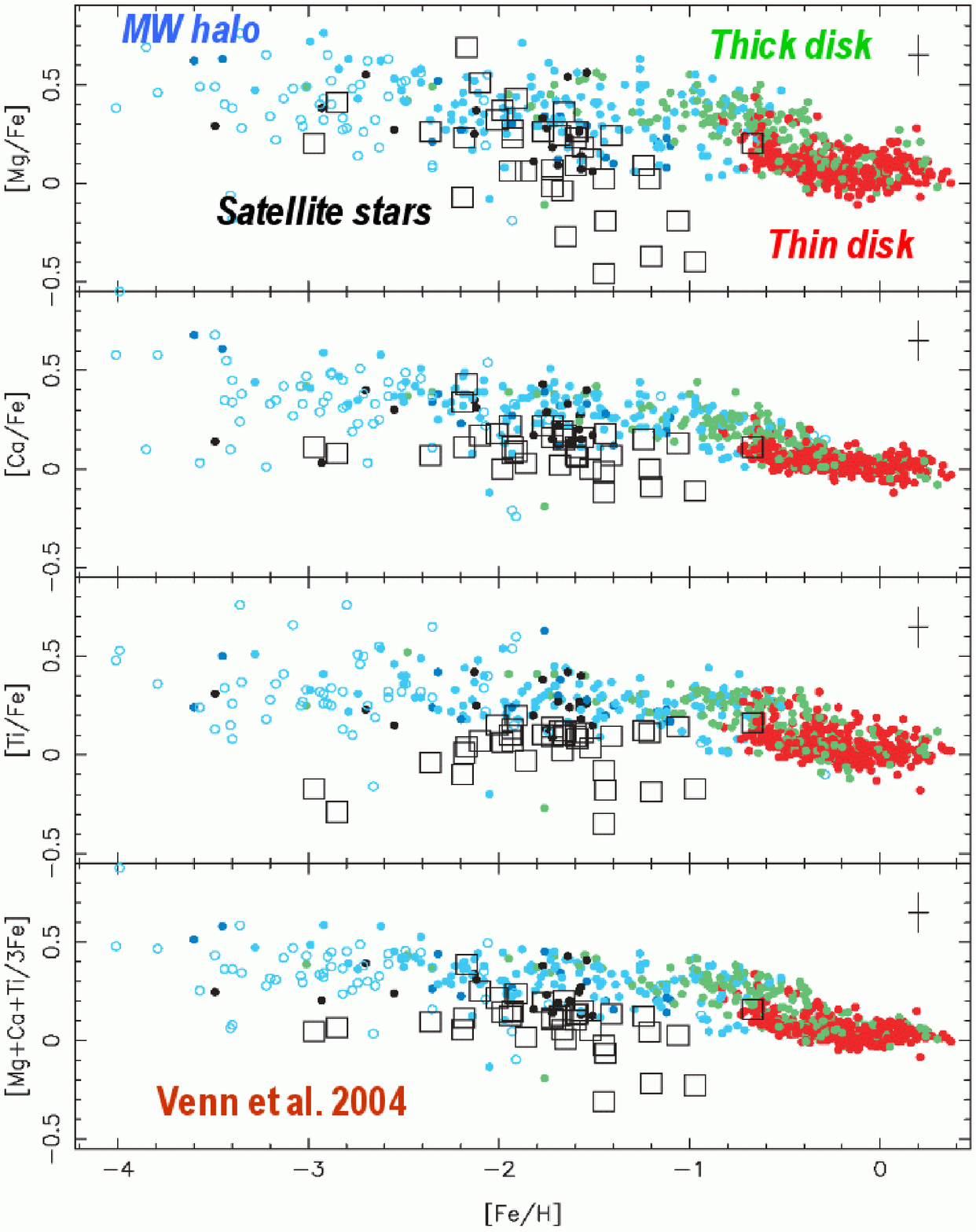}
\qquad
\includegraphics[width=0.47\textwidth]{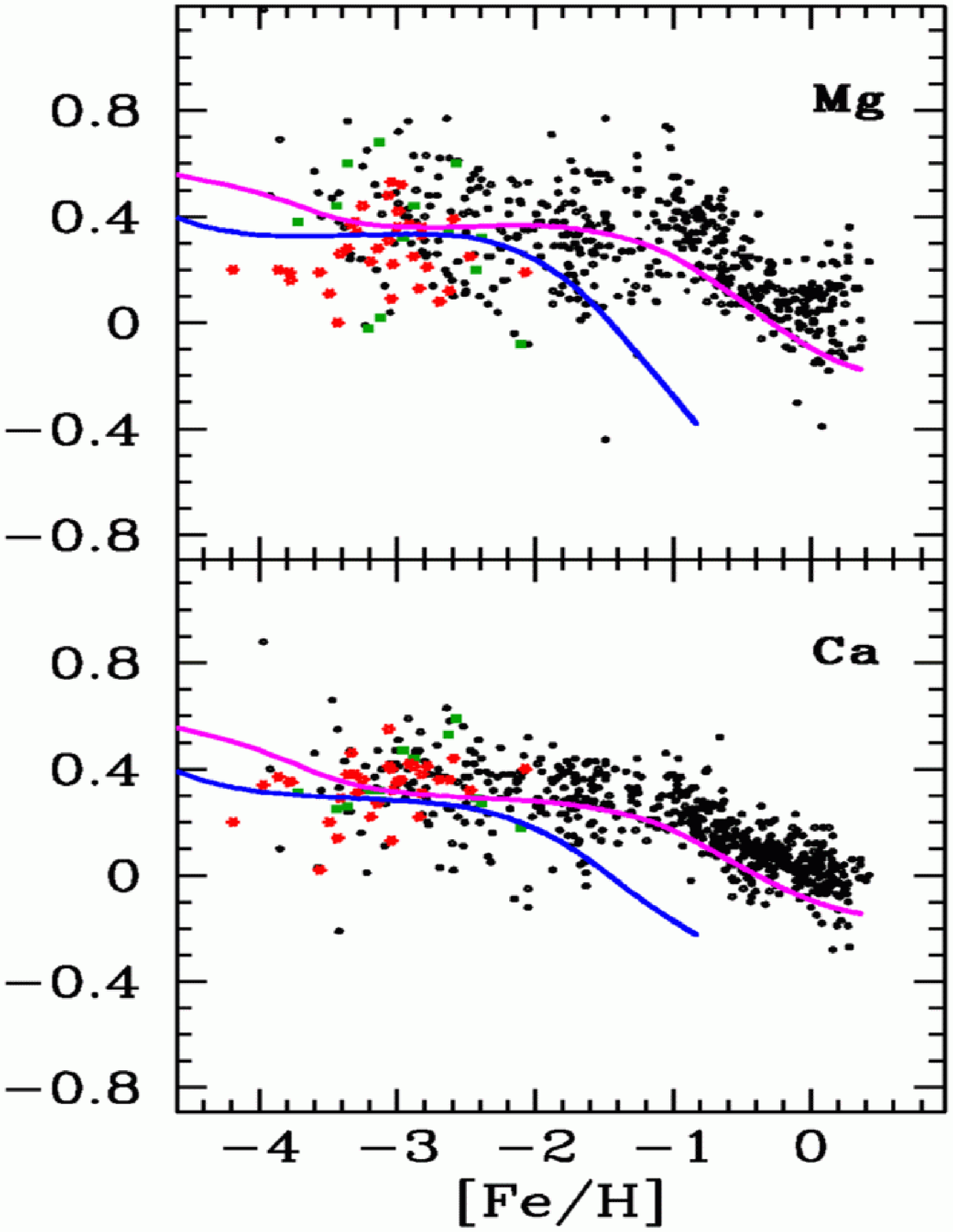}
\caption{{\it Left:} $\alpha$/Fe vs Fe/H relation for various populations of the Milky Way (colour coded) and stars of nearby dwarf galaxies (squares), from Venn et al. (2004). {\it Right:} Observations of $\alpha$/Fe vs Fe/H for stars in the Milky Way disk. The upper curves (reproducing rather well the data) are obtained in a model for the Milky Way, while the lower ones are obtained in  a model with smaller star formation efficiency and mass loss (reproducing a ``typical'' dwarf galaxy), which evolved for 8 Gyr.}
\label{fig:3}       
\end{figure}

Note, however, that abundance patterns vary considerably among Local Group galaxies, calling for more complex explanations than those based simply on differences in timescales.
For instance, while the signature of SNIa (a decrease of $\alpha$/Fe with metallicity) is clearly seen in stars in Sculptor, stars in Draco have uniformly low $\alpha$/Fe ($\sim$solar) for metallicities varying by a factor of $\sim$30 (Fig. 7 in Venn et al. 2004). It is hard to understand why the abundance of Fe (presumably dominated by SNIa, in view of the low $\alpha$/Fe ratio) increases by  a factor of 30 in Draco, while at the same time the $\alpha$/Fe=(Mg+Ca+Ti)/Fe ratio (dominated by Mg, an element not produced in SNIa) remains constant.

The relationship between the abundance patterns of the MW halo and Local Group satellites is addressed in several recent models of galaxy evolution, based on the hierarchical merging framework. Although the various baryonic processes (gaz accretion rate, feedback and impact on outflow rates of various chemical elements) have to be parametrized at present, the everincreasing number of observational data constrains more and more the possible solutions. Robertson et al. (2005) modelled a typical dwarf  spheroidal, a dwarf  irregular and a massive progenitor of the MW halo. Their "gastrophysical" receipes are adapted as to reproduce (among others) the metallicity-luminosity relationship of dwarf galaxies (Fig. 1 b). The $\alpha$/Fe ratio is reproduced well for the dwarf irregular, moderately well for the massive halo progenitor and rather poorly for the dwarf spheroidal galaxy. The constancy of $\alpha$/Fe in Draco is not reproduced or addressed in that work. The early abundance patterns of $\alpha$/Fe do not help distinguishing between dwarf irregular and dwarf spheroidal progenitors as building blocks of the MW halo, but Robertson et al. (2005)  favour the accretion of progenitors of dwarf irregulars on the basis of dynamical arguments (statistics of dark matter halo accretion histories).

Thus, while the abundance patterns of present day dwarf galaxies undoubtely constitute a rich and fascinating subject on its own (see, e.g. Lanfranchi eand Matteucci 2004), it is not clear whether they can teach us directly something about the formation of the MW halo and   hierarchical galaxy formation in general.

\section{The evolution of the Milky Way disk}

The evolution of thin galaxian disks, like the one of the MW, most probably proceeded through smooth accretion of intergalactic gas or minor merger episodes (recent major mergers are excluded on the basis of dynamical arguments, e.g. Toth and Ostriker 1992). The cosmological framework plays then a less important role than in the case of the MW halo formation. However, even within the simple (mostly gas) accretion framework (e.g. Boissier and Prantzos 1999,  Naab and Ostriker 2005) important uncertainties still remain.

\begin{figure}
\centering
\includegraphics[angle=-90,width=0.95\textwidth]{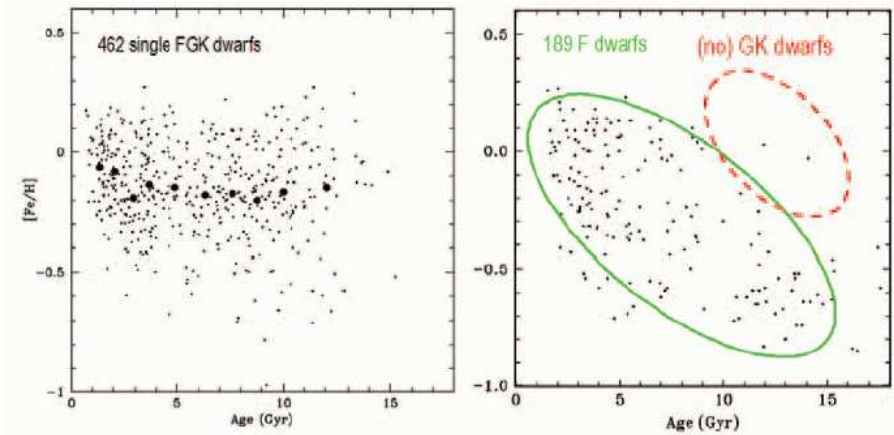}
\caption{Age-metallicity relationship in the solar neighborhood. {\it Left}: data from Nordstrom et al. (2004) for FGK stars within 40 pc. Thick dots indicate average metallicities in the corresponding age bins; no age-metallicity relation appears in those data. {\it Right:} data for 189 F dwarfs from the study of Edvardsson et al. (1993), suggesting an age-metallicity relation; by construction, however, old and metal-rich stars are excluded from that sample. Figure from Nordstrom et al. (2004).}
\label{fig:3}       
\end{figure}

\begin{figure}
\includegraphics[angle=-90,width=0.46\textwidth]{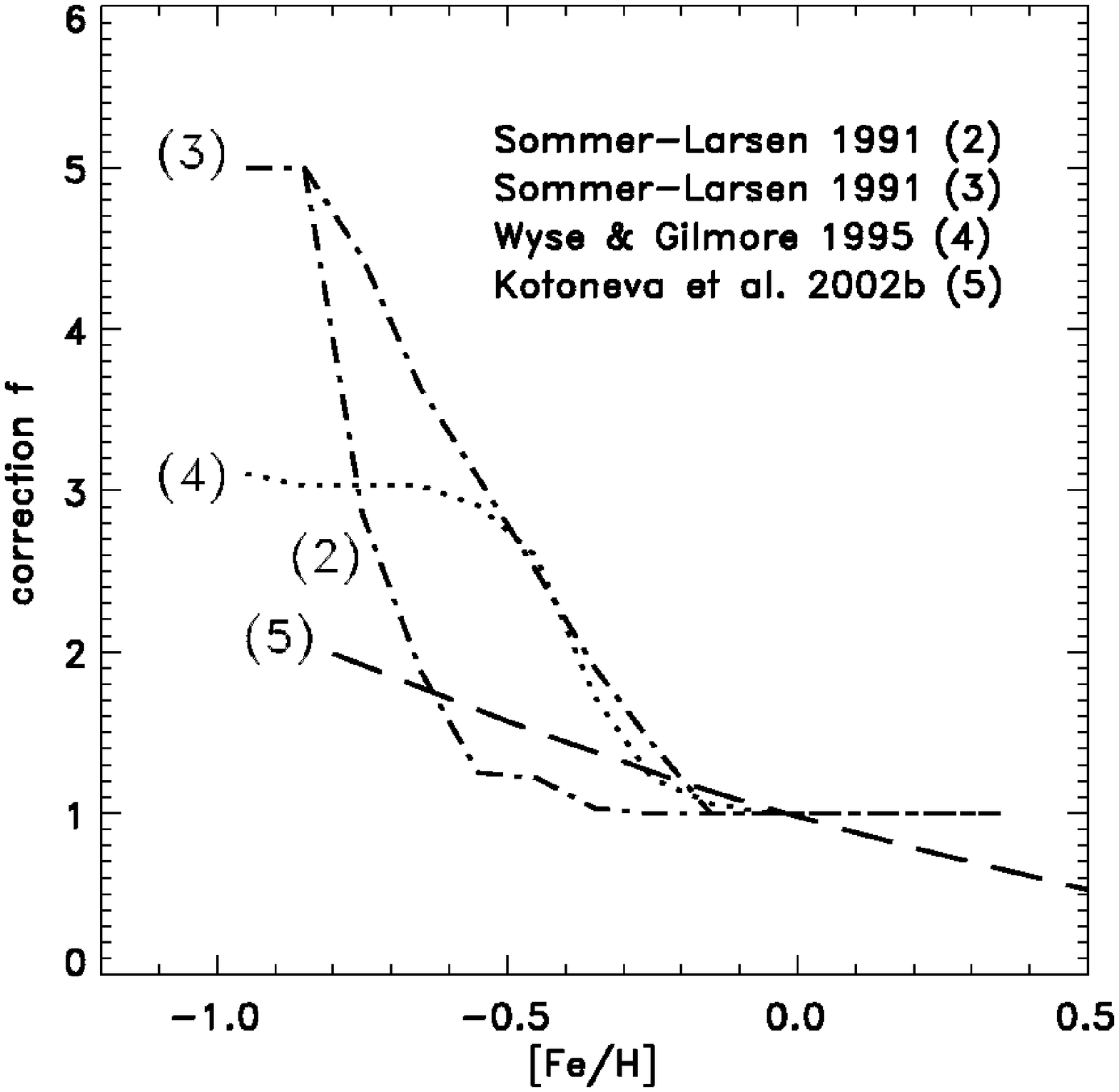}
\qquad
\includegraphics[angle=-90,width=0.46\textwidth]{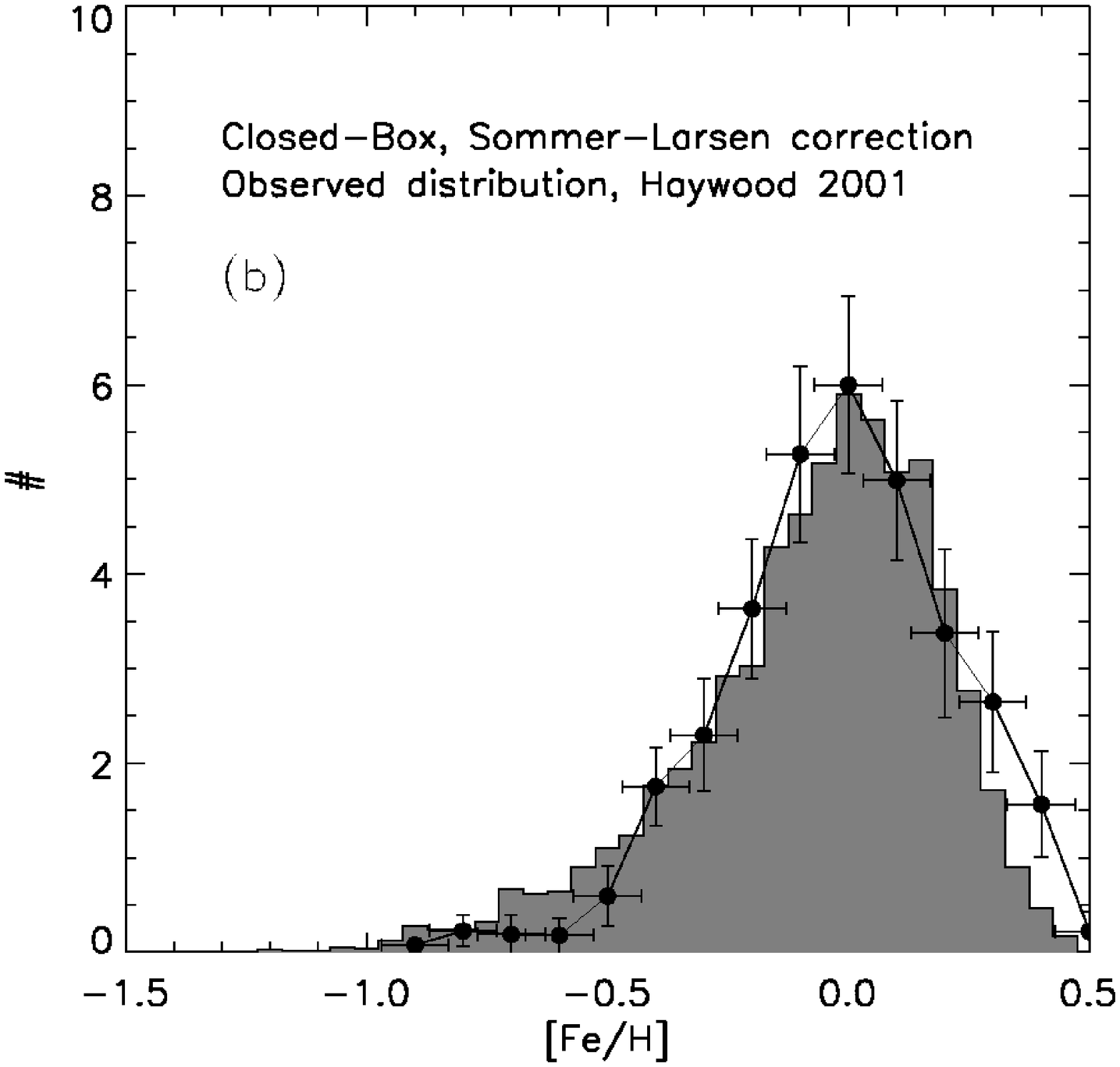}
\caption{{\it Left:} Correcting factors as a function of metallicity, to be applied to the local (thin disk) metallicity distribution, according to Haywood (2006); they are based on estimates of velocity dispersion as a function of metallicity. {\it Right:} If (the inverse of) those correction factors are applied to the results of a closed box model, the resulting metallicity distribution (grey shaded aerea) ressembles closely the observed one (from Haywood 2006).}
\label{fig:3}       
\end{figure}

\subsection{The uncertain evolution of the solar neighborhood}

In studies of the Milky Way evolution (and of GCE in general)  the solar neighborhood plays a pivotal role: the number of available observational data is larger than for any other galactic system, allowing to constrain substantially the number and values of the parameters involved (e.g. Goswami and Prantzos 2000). Among those constraints, the age-metallicity relationship (AMR) and the G-dwarf metallicity distribution are the most important ones.

Stellar ages are much harder to evaluate than stellar metallicities, and the form of the local AMR has varied considerably over the years. The seminal work of Edvardsson et al. (1993) on 189 F-dwarfs established  a clear trend of decreasing metallicity with age, albeit with substantial scatter (Fig. 5 b). 
Such a trend is compatible with (and predicted by) all simple models of local GCE, either closed or open (i.e. with infall) models. It should be noted, however, that the adopted selection criteria in that paper introduced a bias against old metal-rich and young metal-poor  stars.

The large survey  of Nordstrom et al. (2004), concerning $\sim$14000 F and G stars with  3D kinematic information (but less precise spectroscopy than the Edvardsson et al. study), provides a radically different picture: the volume limited subsample of 462 stars with "well-defined" ages withing 40 pc displays a flat AMR (an average metallicity of [Fe/H]$\sim$-0.2 at all ages) with a very large scatter. Acounting for the fact that the oldest stars have the largest age uncertainties does not modify the flatness of the AMR (see also Feltzing et al. 2001).
If confirmed, a flat AMR would require different assumptions than in current models (e.g. substantial late infall to dilute metals).

The form of the second "pillar" of local GCE, namely the local MD, has also been revisited recently by Haywood (2006). He argues that, since the  more metal-poor stellar populations have larger dispersion velocities vertically to the disk,  larger scaleheight corrections should be applied to their numbers in order to get their true surface density. By adopting such steeply decreasing correction factors with metallicity (Fig.  6 a) he finds then that a closed box model nicely account for the corrected MD. He notes that the (volume corrected) local population at -1$<$[Fe/H]$<$-0.5 comprises 15-20 \% of the total; it should then be ``naturally'' considered as the thick disk population, which weights $\sim$7 M$_{\odot}$ pc$^{-2}$, compared to  $\sim$35 M$_{\odot}$ pc$^{-2}$ for the total stellar population in the solar cylinder (Flynn et al. 2006).

The relationship of the thick and thin disks is not yeat clear (see Sec. 3.2); if thick disk stars were mostly accreted from merging/disrupted satellites (e.g. Abadi et al. 2003), that population cannot be considered as belonging to the early phase of the disk and the arguments of Haywood (2006) do not hold. Also, in view of the large scaleheight of the thick disk (1400 pc)  and of  the elliptical orbits of its stars (vs. circular orbits for those of the thin disk) one may wonder how large the local ``chemical box'' could be and still be considered as a system with uniform composition. It is clear, however, that corrections to the observed local MD should be carefully considered before comparing to GCE models and this would certainly impact on the required infall timescales (which could be smaller than the ``canonical'' value of $\sim$7 Gyr currently adopted).

\subsection{The thick disk}

More than 20 years after its recognition as a separate sub-component of the MW (Gilmore and Reid 1983), the properties and origin of the thick disk are still a matter of debate. Recent surveys (Subiran and Girard 2005, Reddy et al. 2006) established clearly differences in ages, metallicities  and abundance patterns, the thick disk being a few Gyr older than the thin (Fig. 7 a), less metallic by $\sim$-0.5 dex and having slightly larger $\alpha$/Fe ratios.

Each of the scenarios proposed for the thick disk encounters some problems (see Dalcanton et al. 2005 and Reddy et al. 2006 for discussion): (a)
in view of their different kinematic properties, it appears difficult to explain the thick disk as the vertically ``heated'' (by mergers)  early thin disk; (b) accretion of stars from disrupted satellite galaxies (Abadi et al. 2003) can hardly account for the observed regular abundance patterns and their small dispersion; (c) finally, formation during an early intense period of gas-rich mergers, appears promising, with the thin disk formed a few Gyr later from the gas of the mergers plus infalling pristine gas (which dilutes the metallicity of the merger gas, so that the thin disk starts with a smaller metallicity than the maximal one reached by the thick disk; Fig. 7b).

Scenario (c) is developed in chemodynamical simulations of Brook et al. (2005) and reproduces quite satisfactorily the abundance patterns of $\alpha$/Fe for the halo, thick and thin disks (Fig. 8 a); however, as noticed in Dalcanton et al. (2005), it produces a thick disk more compact than the thin disk (scalelength ratio smaller than unity), whereas the opposite is observed in the MW and most disk galaxies (Fig. 8 b).

\begin{figure}
\includegraphics[angle=-90,width=0.47\textwidth]{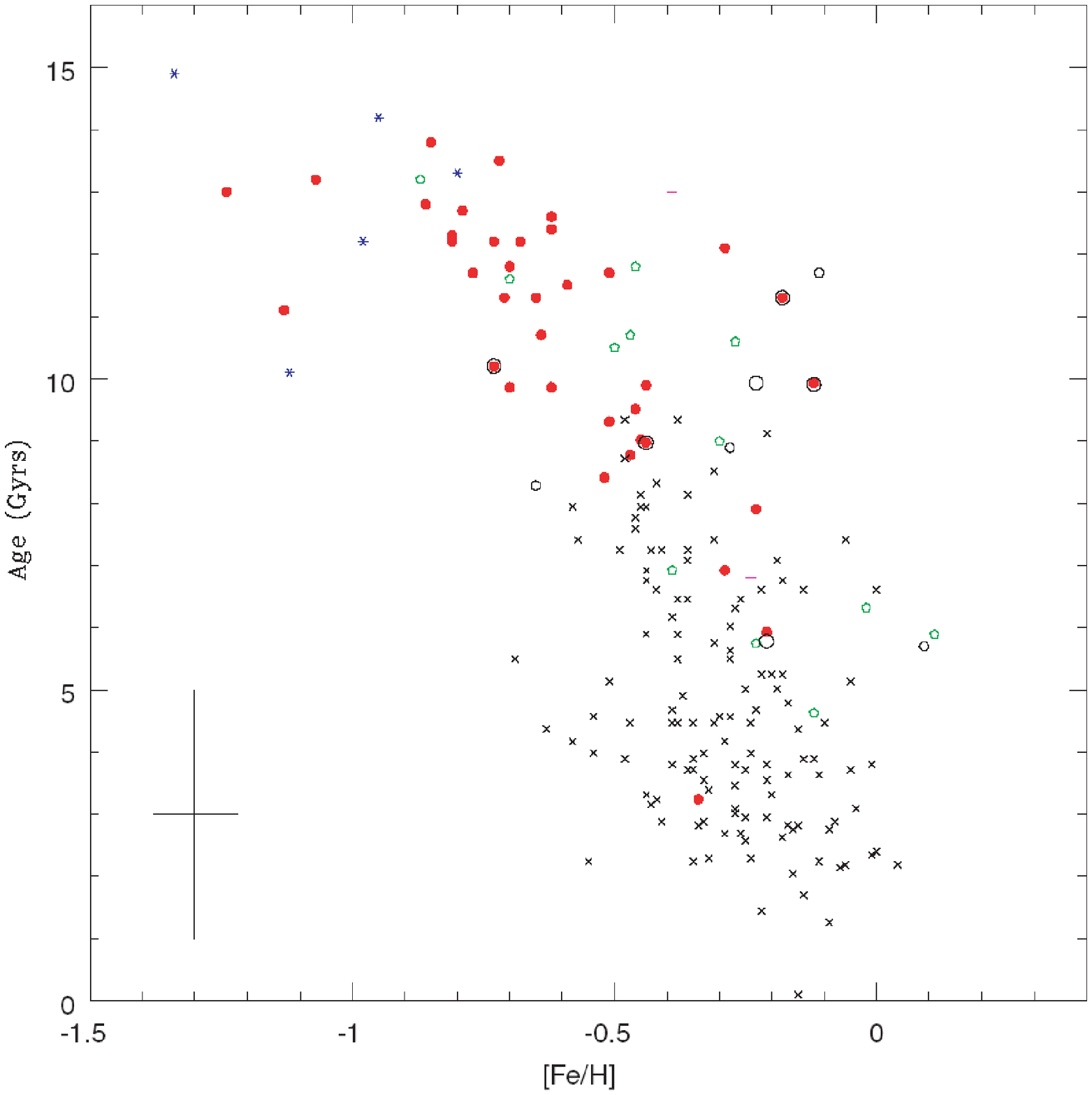}
\qquad
\includegraphics[angle=-90,width=0.49\textwidth]{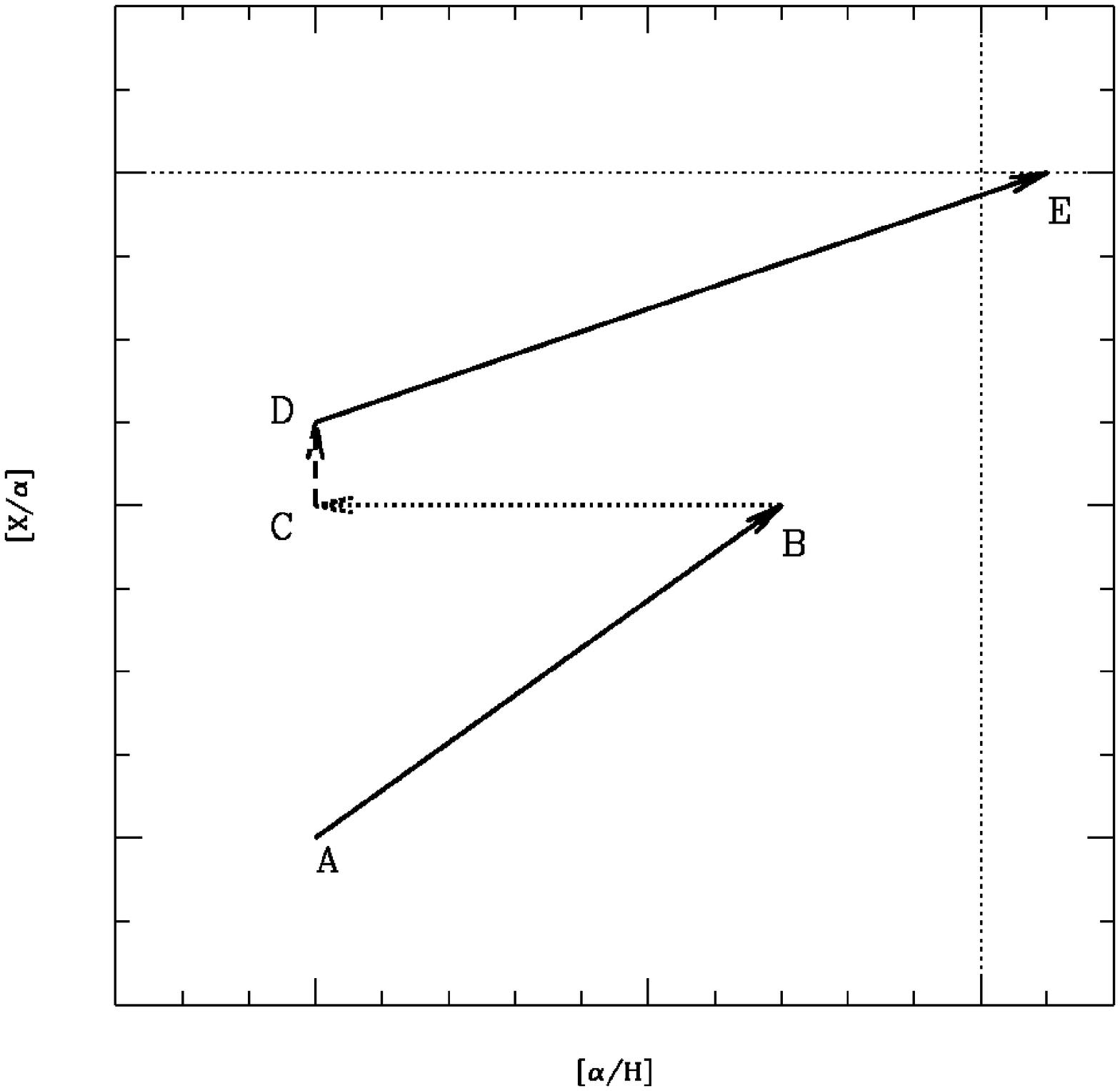}
\caption{{\it Left:} Age-metallicity relation for stars in the halo and thick and thin disks (colour coded) from Reddy et al. (2006). {\it Right:} Schematic evolution of the abundance ratio X/$\alpha$, where X is a secondary element, in the thick disk (AB), the quiescent (no SF and pristine infall) period (BC) and the thin disk (DE), according to Reddy et al. (2006).}
\label{fig:3}       
\end{figure}

\begin{figure}
\includegraphics[angle=-90,width=0.46\textwidth]{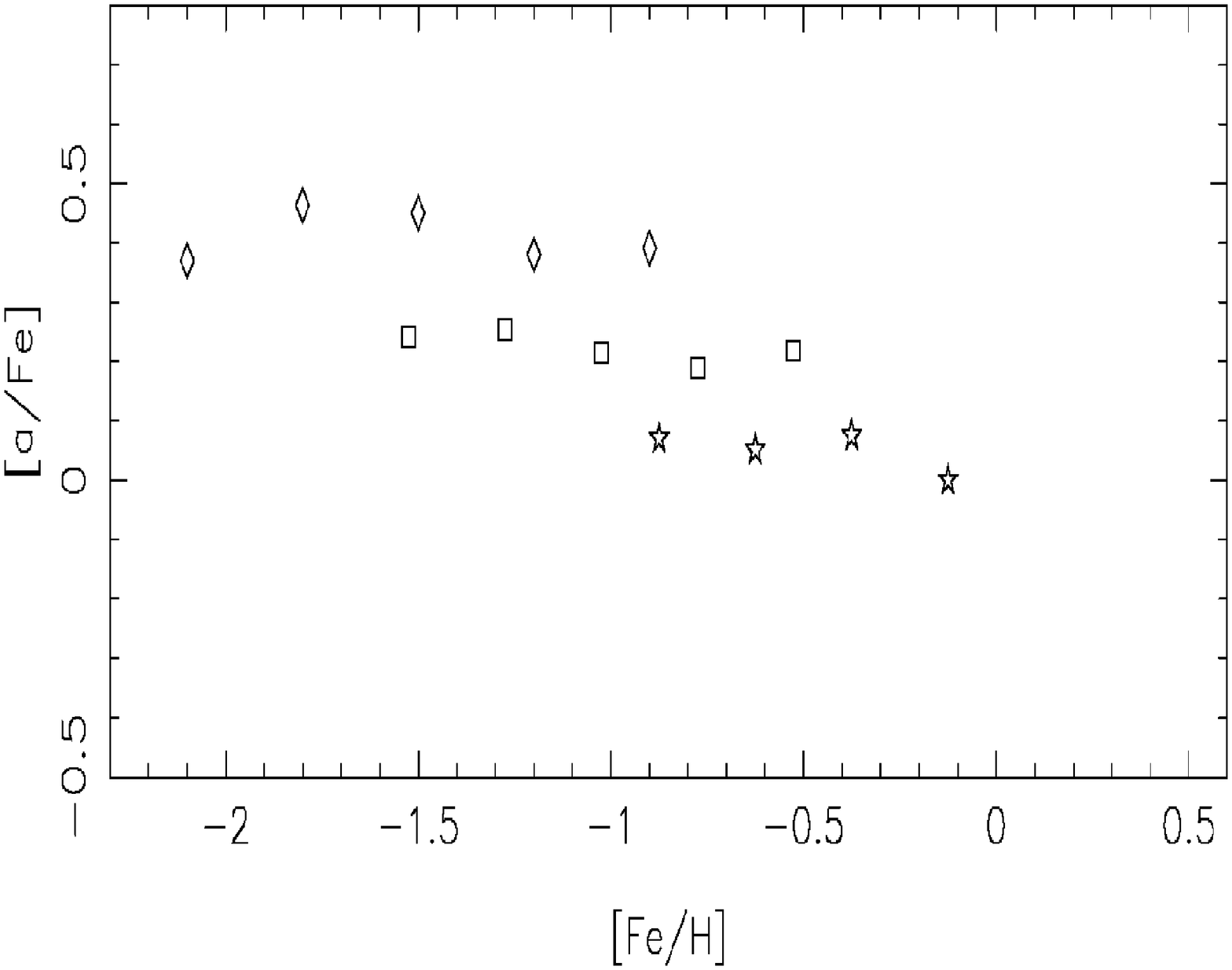}
\qquad
\includegraphics[angle=-90,width=0.46\textwidth]{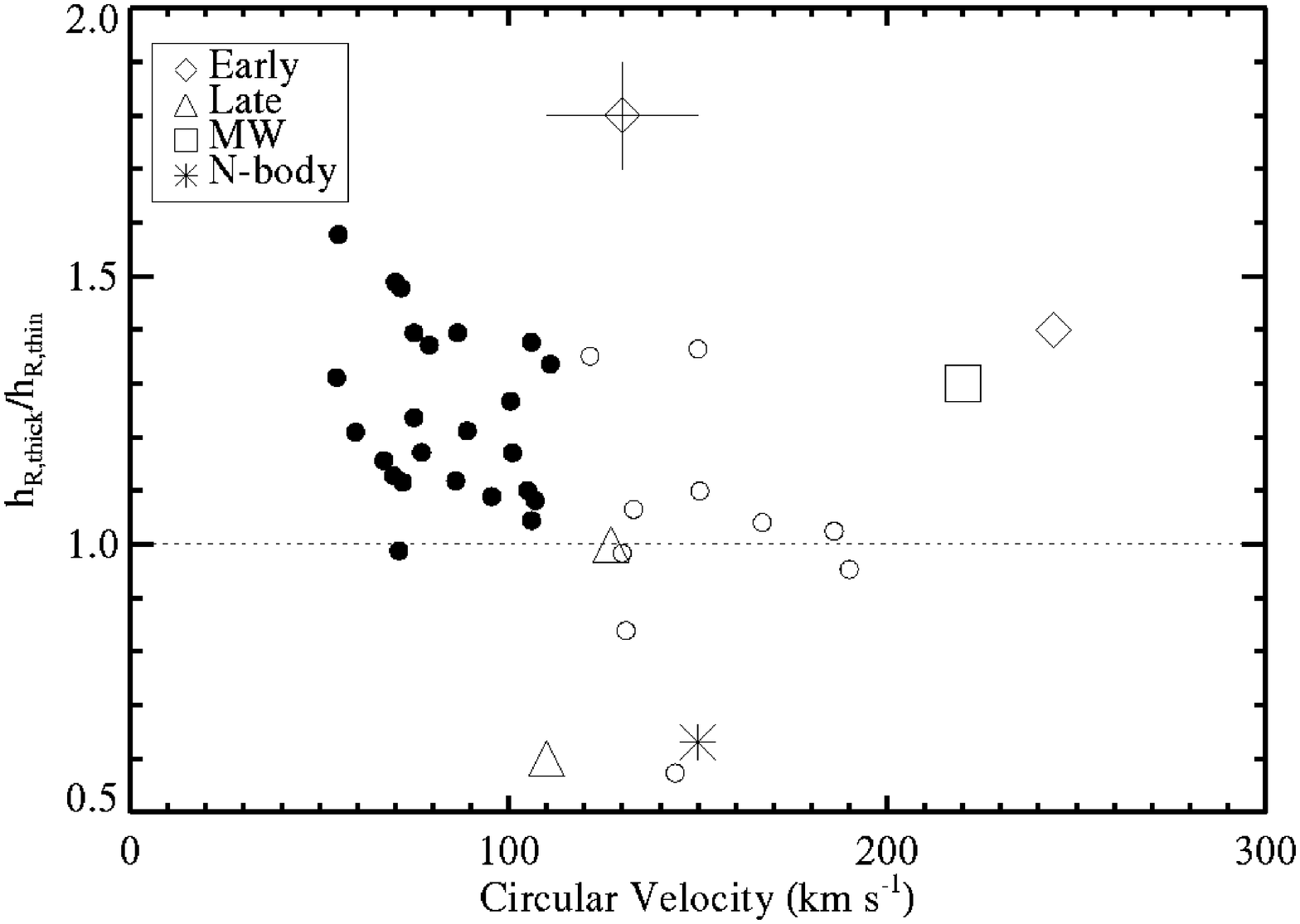}
\caption{
{\it Left} panel: $ \alpha$/Fe ratio evolution in the halo (triangles), thick disk (squares) and thin disk (asterisks), from models of Brook et al. 
(2005). {\it Right}: Ratio of scalelengths thick/thin disks vs circular velocity (mass) of the galaxy; the MW is indicated with a large open square and N-body results for the MW with an asterisk (from Dalcanton  2005).
}
\label{fig:3}       
\end{figure}

\subsection{Abundance gradients in the disk}
 
Abundance gradients are a well established observed feature of galactic disks, and they result naturally in inside-out disk formations schemes. Their magnitude and evolution are shaped by various factors, both local (SFR, infall/outflow rates) and non-local ones (e.g. strength of galactic bars, dynamicaly mixing inwards gas from the periphery and reducing the gradients). 

In the case of the MW disk, the magnitude of the abundace gradient (depending on adopted tracers and analysis methods) is not well established yet, with values of dlog(O/H)/dR varying between -0.07 dex/kpc and -0.04 dex/kpc (e.g. Daflon and Kunha 2004, Rudolph et al. 2006). Large (absolute) values suggest a minor role for the Galactic bar and  a strong radial dependence of the SFR/infall rate.

The evolution of the abundance gradient is even more difficult to measure, because of large uncertainties in age determinations. Maciel et al. (2005, 2006), using observations of planetary nebulae of various ages and other tracers, estimate that the abundance gradient was about twice as large $\sim$7 Gyr ago, in agreement with predictions of GCE models by Hou et al. (2000, Fig. 9 a); such an evolution, if real, would also minimize the role of the Galactic bar throughout most of the disk history.

The MW appears to be a typical disk galaxy as far as its abundance gradient is concerned. Observations show that those gradients (in dex/kpc) are anticorrelated with disk size (Fig. 9 b), suggesting some kind of ``homologuous disk evolution''; only simple disk models (with the disks growing in dark matter haloes) have reproduced that relation up to now (Prantzos and Boissier 2000).

\begin{figure}
\centering
\includegraphics[angle=-90,width=0.46\textwidth]{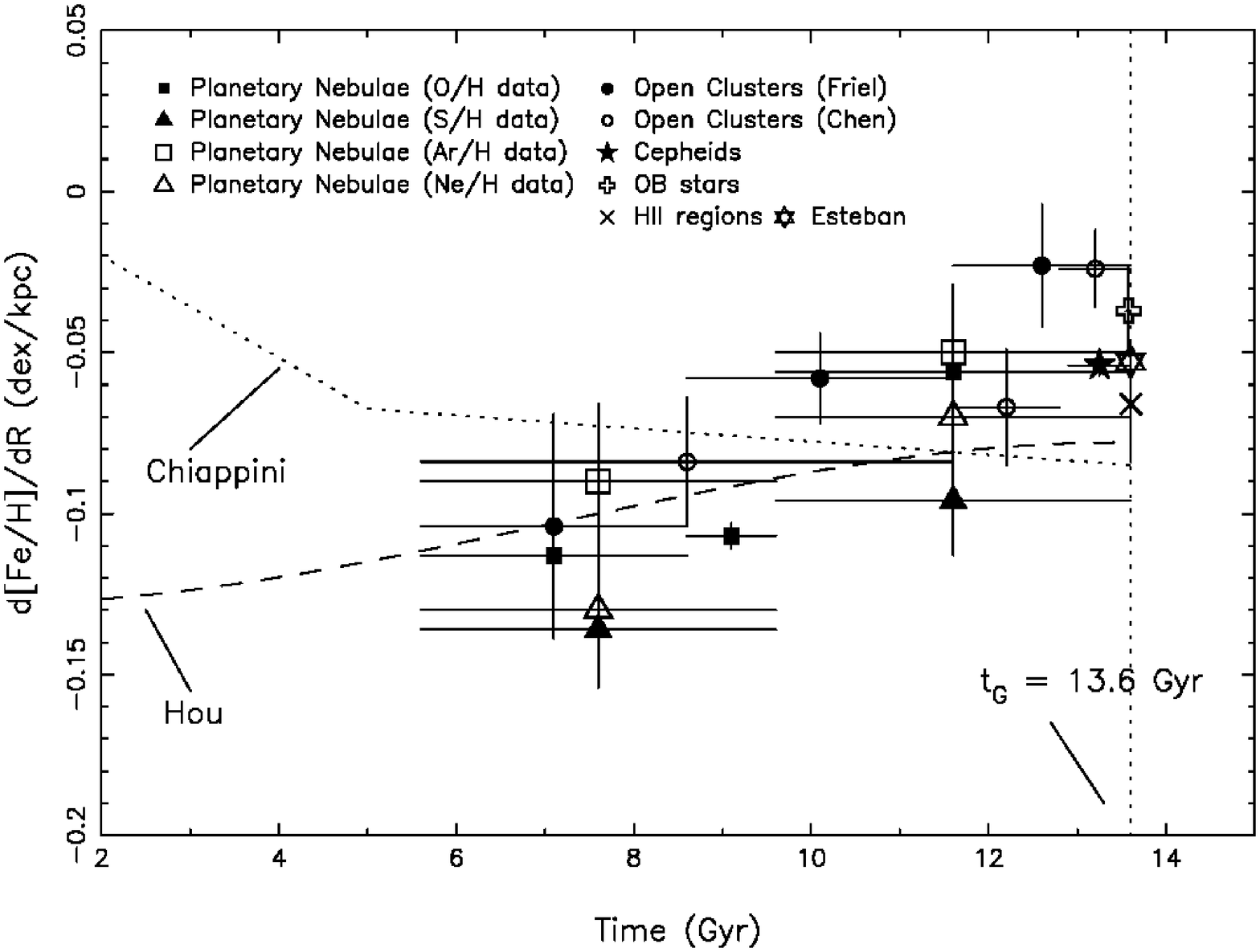}
\qquad
\includegraphics[angle=-90,width=0.46\textwidth]{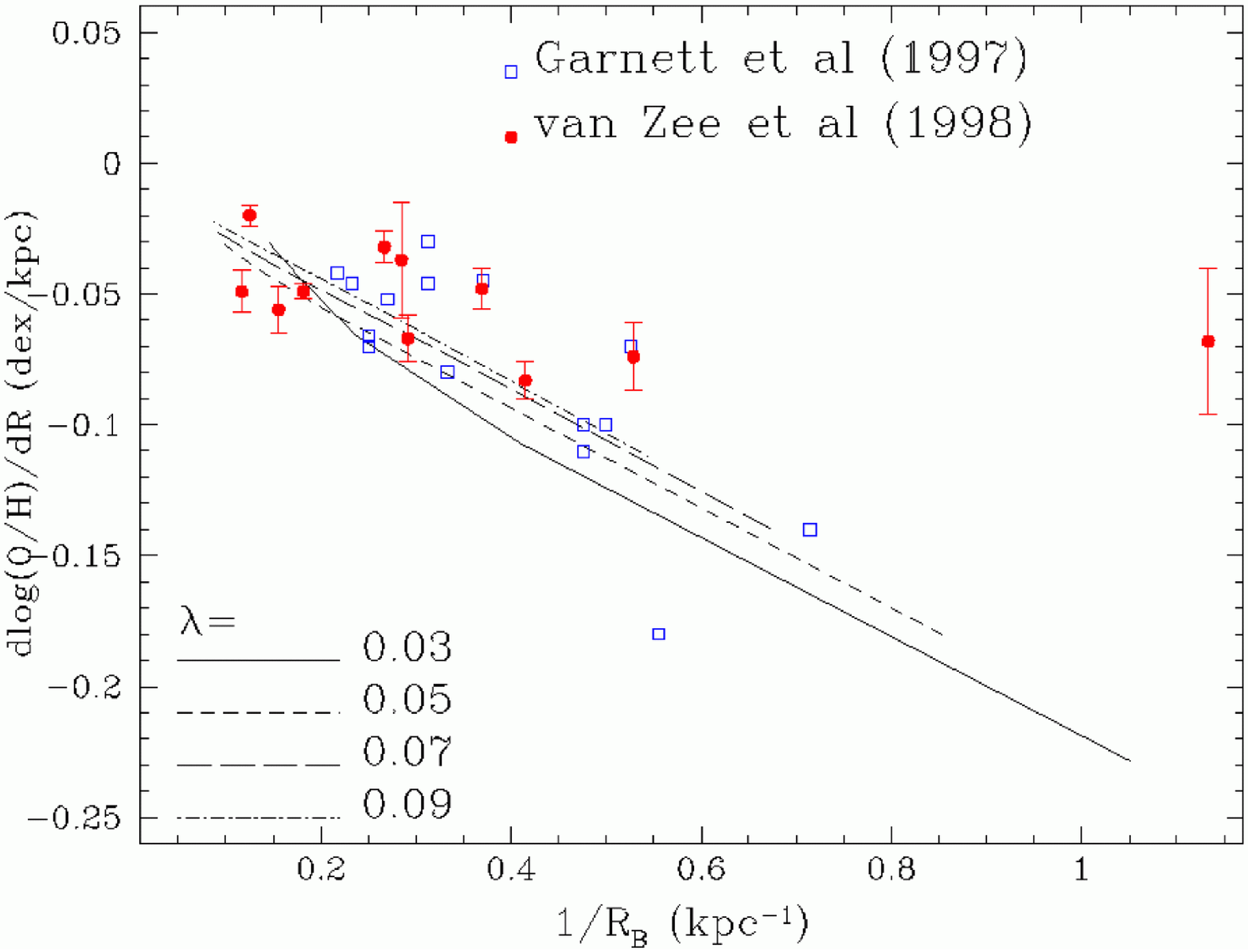}
\caption{{\it Left:} Estimates of the evolution of the Fe abundance gradient in the Milky Way from various tracers; they are compared to theoretical predictions (from Maciel et al. 2006). {\it Right}: Abundance gradients for disk galaxies vs the inverse of the scalelength in the B-band, compared to model results (from Prantzos and Boissier 2000). }
\label{fig:3}       
\end{figure}

\bigskip

{\small {\bf Acknowledgement:} I am grateful to the organisers for their kind invitation in such an interesting meeting and for financial support.}

\end{document}